\documentclass[prc,preprint,showpacs,superscriptaddress,floatfix,amsmath,amssymb,nofootinbib]{revtex4}
\usepackage{graphicx}% Include figure files
\usepackage{color}
\usepackage{CJK}
\usepackage{ulem}

\newcommand{\beq}{\begin{equation}}
\newcommand{\eeq}{\end{equation}}
\newcommand{\beqn}{\begin{eqnarray}}
\newcommand{\eeqn}{\end{eqnarray}}

\newcommand{\gev}{\mathrm{GeV}}
\newcommand{\GeV}{\mathrm{GeV}}

\newcommand{\fm}{\mathrm{fm}}
\newcommand{\m}{\mathrm{M}}
\newcommand{\M}{\mathrm{M}}
\newcommand{\km}{\mathrm{km}}

\begin{document}
%\begin{CJK*}{GBK}{song}
%\CJKindent
\title{The maximum mass of dark matter existing in compact stars based on the self-interacting fermionic  model}
\author{X. D. Wang}%\thanks{e-mail: bqi@sdu.edu.cn}
\author{B. Qi}\thanks{e-mail: bqi@sdu.edu.cn}
\author{N. B. Zhang}
\author{S. Y. Wang}
\affiliation{Shandong Provincial Key Laboratory of Optical Astronomy
and Solar-Terrestrial Environment, School of Space Science and Physics, Institute of Space Sciences, Shandong University, Weihai, 264209, People's
Republic of China}

\date{\today}
%\received{(received date)} \revised{(revised date)}
%\accepted{(Day Month Year)}
%\comby{(xxxxxxxxxx)}

\begin{abstract}
By assuming that only gravitation acts between dark matter (DM) and normal matter (NM), we studied DM admixed neutron stars (DANSs) using the two-fluid TOV equations.
The NM and DM of compact stars are simulated by the relativistic mean field (RMF) theory and non-self-annihilating self-interacting fermionic model, respectively.
The effects of the particle mass of fermionic DM $m_f$ and the interaction strength parameter $y$  on the properties of DANSs are investigated in detail.  $m_f$ and $y$ are considered  as the free parameters due to the lack of information about the particle nature of DM so far.
For a DANS,  we suggest a simple universal relationship  $ M_D^{\max}=(0.267 y +0.627-3.21\frac{M_N}{\M_{\odot}})( \frac{1\GeV}{{m_f}})^2 \M_{\odot} $ for  $y>100$, where $ M_D^{\max} $ is the maximum mass of DM existing in DANSs  and $M_N$ is the mass of the neutron star without DM. For free fermion DM model ($y$=0),  the relationship becomes $ M_D^{\max}=(0.627-0.027\frac{M_N^2}{\M_{\odot}^2}) ( \frac{1\GeV}{{m_f}})^2 \M_{\odot}$.
The radius of DM $R_D$ shows a linear relationship with $M_D^{\max}$ in DANSs, namely $R_D=(7.02 \frac{M_D^{\max}}{ \M_{\odot}}+1.36)$~km.
These conclusions are independent of the different NM EOSs from RMF theory.  Such a kind of universal relationship  connecting the nature of DM particle and mass of stars  might shed light on the constraining the nature of the DM by indirect method.

\end{abstract}

%\keywords: neutron stars, dark matter,  universal relationship

\pacs{95.35.+d, 97.60.Jd, 26.60.-c, 21.60.Jz }
%95.35.+d	Dark matter (stellar, interstellar, galactic, and cosmological)
%97.60.Jd	Neutron stars
%26.60.-c	Nuclear matter aspects of neutron stars	
%21.60.Jz   Nuclear Density Functional Theory and extensions (includes Hartree-Fock and random-phase approximations)
\maketitle

\section{Introduction}
Nowadays, the existence of dark matter (DM) has been well accepted, and the observations expose that
most of the mass of the Universe is in the form of DM\cite{zwickyAJ1937,BetouleAA2014,AdeAA2014}.
There are many suggested DM candidates, such as neutrinos, weakly interacting sub-eV particles (WISPs) and weakly interacting massive particles (WIMPs)\cite{ NarainPRD2006,KouvarisPRD2015,EbyJHEP2016, KlasenPPNP2015}.
However, the nature of DM, including the mass of particle and interactions, is still the mystery. Thus, constraining the nature of DM through direct or indirect methods  becomes a very hot topic in both astrophysics and particle physics~\cite{ KlasenPPNP2015, FengARAA2010}.
There are three main ways to detect DM particles: using particle accelerators to find the possible candidates for DM\cite{CarpenterPRD2013,AadJHEP2013,ChatrchyanJHEP2012},
detecting the signal of DM particle annihilation in the galactic halo\cite{FengAIP2013}, and the signals of DM-nucleus scatterings in terrestrial detectors\cite{CuiPRL2017}.
By now, the latest experimental results are not conclusive.

On the other hand,  indirect method by studying the DM effects on compact stars such as neutron stars \cite{SandinAP2009} and quark stars\cite{AngelesPRL10,MukhopadhyayPRD2016}, has
obtained attention in recent years. The general effect induced by DM inside neutron star is complicated due to the lack of information about the particle nature of DM.
Therefore, it is of great significance to study the potential effects of DM on the properties of neutron stars. %The most studied hypothesis is that DM is made of WIMPs.
%Non-self annihilating DM  would simply accumulate inside NS  and affect the stellar structure.
Recent studies have been done to
explore compact stars with non-self-annihilating fermionic dark
matter to analyze the gravitational effects of DM
on the stellar matter using the two-fluid TOV formalism (see, e.g., Refs.~\cite{LeungPRD2011,LiAP2012,XiangPRC2014,TolosPRD2015}).
Ref.~\cite{LeungPRD2011} suggested a new class of compact stars which consists of a small NM core embedded in a DM halo  when considering DM particles of mass about 1 GeV. Ref.~\cite{LiAP2012} found that DM inside the star would soften the equation of state more strongly than that
of hyperons, and reduce largely the maximum mass of the star.
In Ref.~\cite{XiangPRC2014}, it is found that the mass-radius relationship of the DM admixed neutron stars (DANSs) depends sensitively on
the mass of DM candidates, the amount of DM, and interactions among DM candidates.  Ref.~\cite{TolosPRD2015} suggested the dark compact planets with Earth-like
masses or Jupiter-like masses.

In this paper, we focus on exploring a simple formula connecting the microcosmic nature of DM particle and its macrocosmic mass existing in DM admixed neutron stars (DANSs). Ref.~\cite{NarainPRD2006} had ever demonstrated that for a pure fermionic DM star, there is a simple universal relationship  $ M_D^{\max}=(0.269 y+0.627)( \frac{1\GeV}{{m_f}})^2 \M_{\odot}$, where $ M_D^{\max} $ is the maximum mass of compact star,  $ m_f$ is particle mass of fermionic DM, $y$ is interaction strength parameter between DM particles.
Following their work, based on the non-self-annihilating self-interacting fermionic DM model,  we explore the possible relationship between the  maximum mass of DM existing in DANSs and the properties of DM. Searching for such a universal relationship  might shed light on the constraining the nature of the DM by indirect method.

The paper is organized as follows: In Sec.~II, we briefly discuss the main theory used in this paper, include the  self-interacting fermionic DM model,  relativistic mean field (RMF) theory  and the two-fluid TOV equations. In Sec.~III, the effects of particle mass of fermionic DM $m_f$ and the interaction strength parameter $y$ on the properties of DANSs are investigated in detail.
The relationship between the  maximum mass of DM existing in DANSs and the properties of DM are studied.
Finally we summarize our work in  Sec.~IV.

\section{Formalism}
%\subsection{The equation of state for a gas of free fermions}
In this paper,  we use the non-self-annihilating self-interacting fermionic model to simulate DM in DANSs, where the detailed formulism can be seen in Ref.~\cite{MukhopadhyayPRD2016}. We only show the energy density and pressure here:

\begin{eqnarray}
\varepsilon&=&\frac{1}{\pi^2}\int_{0}^{k_F}k^2\sqrt{m_f^2+k^2}dk+\left[\left(\frac{1}{3\pi^2}\right)^2y^2z^6\right]\nonumber\\
&=&\frac{m_f^4}{8\pi^2}\left[(2z^3+z)\sqrt{z+z^2}-\sinh^{-1}(z)\right]+\left[\left(\frac{1}{3\pi^2}\right)^2y^2z^6\right],
\end{eqnarray}
\begin{eqnarray}
p&=&\frac{1}{3\pi^2}\int_{0}^{k_F}\frac{k^4}{\sqrt{m_f^2+k^2}}dk+\left[\left(\frac{1}{3\pi^2}\right)^2y^2z^6\right]\nonumber\\
&=&\frac{m_f^4}{24\pi^2}\left[(2z^3-3z)\sqrt{z+z^2}+3\sinh^{-1}(z)\right]+\left[\left(\frac{1}{3\pi^2}\right)^2y^2z^6\right],
\end{eqnarray}
where $m_f$ is the particle mass of fermionic DM, $k$ is the momentum, $z=k_F/m_f$ is the dimensionless Fermi
momentum and $y$ is the dimensionless interaction strength
parameter, which is defined as $ y = m_f/m_I$ (the interaction
mass scale $m_I$)~\cite{NarainPRD2006,MukhopadhyayPRD2016}.  For weak interaction, the typical scale is $m_I$$\sim300$ GeV,  as the
expected masses of W or Z bosons. For strongly interacting DM particles, $m_I$
is assumed to be $\sim 100$ MeV, according to the gauge theory of the
strong interactions \cite{NarainPRD2006,MukhopadhyayPRD2016}.

For the NM in DANSs, we adopt the RMF theory, which has achieved great success in the description of nuclear matter and finite nuclei in the past several decades \cite{Meng06,Ring96}. Meanwhile, the RMF theory has been used to study the neutron stars and obtained a lot of valuable results \cite{Ban04,Wang14,Sun08,Glendenning98,Jiang12,QiRAA2016,ZhangCPC2017}. The start of the RMF theory is an effective Lagrangian density. In the present work, we use the density dependent RMF theory where the effective Lagrangian density for nuclear matter is written as:
\begin{eqnarray}\label{ld}
{\cal L}&=&\sum_{B}\overline{\psi}_B[i\gamma^{\mu}\partial_{\mu}-m_B-g_{\sigma B}\sigma -g_{\omega B}\gamma^{\mu}\omega_{\mu}
-g_{\rho B}\gamma^{\mu}\tau_B\cdot\rho_{\mu}\nonumber \\
&-&e\gamma^{\mu}A_{\mu}\frac{1-\tau_{3B}}{2}]\psi_B+\frac{1}{2}\partial_{\mu}
\sigma\partial^{\mu}\sigma-\frac{1}{2}m^2_{\sigma}\sigma^2\nonumber \\
&-&\frac{1}{4}\omega_{\mu\nu}\omega^{\mu\nu}+\frac{1}{2}m^2_{\omega}\omega_{\mu}
\omega^{\mu}-\frac{1}{4}\rho_{\mu\nu}\rho^{\mu\nu}+\frac{1}{2}m^2_{\rho}\rho_{\mu}\rho^{\mu}-\frac{1}{4}A_{\mu\nu}A^{\mu\nu}.
\end{eqnarray}
The specific meanings of each parameter will not be introduced in detail here and can be found in Ref.~\cite{Ban04}. The Lagrangian density for the neutron stars is different from the one for nuclear matter as the coulomb field is neglected and an additional term for leptons is needed.
By solving the equations of motion, the corresponding energy density and pressure for the NM are \cite{Ban04}
\begin{eqnarray}
\varepsilon&=&\frac{1}{2}m^2_{\sigma}\sigma^2+\frac{1}{2}m^2_{\omega}\omega_{0}^2+\frac{1}{2}m^2_{\rho}\rho^2_{0,3}+\frac{1}{\pi^2}\sum_{B}\int^{k_B}_0k^2dk\sqrt{k^2+(m_B+g_{\sigma B}\sigma)^2} \nonumber \\
&+&\frac{1}{\pi^2}\sum_{\lambda=e^-,\mu^-}\int_0^{k_{\lambda}}k^2dk\sqrt{k^2+(m_\lambda+g_{\sigma B}\sigma)^2},\\
p&=&-\frac{1}{2}m^2_{\sigma}\sigma^2+\frac{1}{2}m^2_{\omega}\omega_{0}^{2}+\frac{1}{2}m_{\rho}^2\rho^2_{0,3}+\frac{1}{3\pi^2}\sum_{B}\int_{0}^{k_{B}}\frac{k^4}{\sqrt{k^2+(m_B+g_{\sigma B}\sigma)^2}}dk\nonumber\\ &+&\sum_B\rho_B\Sigma^R_{0B}
+\frac{1}{3\pi^2}\sum_{\lambda=e^-,\mu^-}\int_{0}^{k_{\lambda}}dk\frac{k^4}{\sqrt{k^2+m_{\lambda}^2}},
\end{eqnarray}
where $\sum_{\mu B}^{R}$ is the time component of the ``rearrangement'' term.

%\subsection{two fluid tolmann-oppenheimervolkoff equations}
The compact stars which made of DM and NM are inherently two-fluid system. If  NM and DM couple essentially only through gravity, DANSs can be studied by the TOV equations for two-fluid separately
\cite{SandinAP2009,MukhopadhyayPRD2016,LeungPRD2011,TolosPRD2015}:
\begin{eqnarray}
% \nonumber to remove numbering (before each equation)
\frac{dp_1}{dr} &=& -\frac{GM(r)\varepsilon_1(r)}{r^2} \left(1+\frac{p_1(r)}{\varepsilon_1(r)}\right)
\times\left(1+4\pi r^3\frac{p_1(r)+p_2(r)}{M(r)}\right)
\left(1-2G\frac{M(r)}{r}\right)^{-1},\nonumber \\
\frac{dp_2}{dr}&=&-\frac{GM(r)\varepsilon_2(r)}{r^2} \left(1+\frac{p_2(r)}{\varepsilon_2(r)}\right)
\times\left(1+4\pi r^3\frac{p_1(r)+p_2(r)}{M(r)}\right)
\left(1-2G\frac{M(r)}{r}\right)^{-1}, \nonumber \\
\frac{dM_1}{dr}&=&4\pi r^2\varepsilon_1(r),\nonumber \\
\frac{dM_2}{dr}&=&4\pi r^2\varepsilon_2(r), \nonumber \\
M(r)&=&M_1(r)+M_2(r),
\end{eqnarray}
where $M(r)$  represents the total mass at the radius $r$.  $p_1$, $p_2$, $\varepsilon_1$, $\varepsilon_2$ represent the pressure and energy density of DM and NM.

\section{ Discussion}

To discuss the effects of fermionic DM on the properties of DANSs, we consider the particle mass of fermionic DM $m_f$ and  the strength parameter $y$ as the free
parameters  due to the lack of information about the particle nature of DM so far. Here we calculate all the results with  $m_f=1.0$, $ 1.1$, $ 1.3$, $ 1.5$, $ 2$, $ 5$, $ 10$, $ 20$, $ 30$, $ 40$, $ 50$, $ 60$, $ 70$, $ 80$, $ 90$, $ 100~ \gev$ and  $y =0 $, $ 1 $, $ 2 $, $ 3 $, $ 4 $, $ 5 $, $ 50 $, $100$, $ 200$, $300$,  $400$, $500$, $ 600$, $700$, $800$, $ 900$, $1000$. For describing NM in DANSs, different parameter sets DDLN, DDME1, DDME2, PKDD from RMF~\cite{Meng06} are adopted.

\begin{figure}[h!]
    \centering
    \includegraphics[width=12cm]{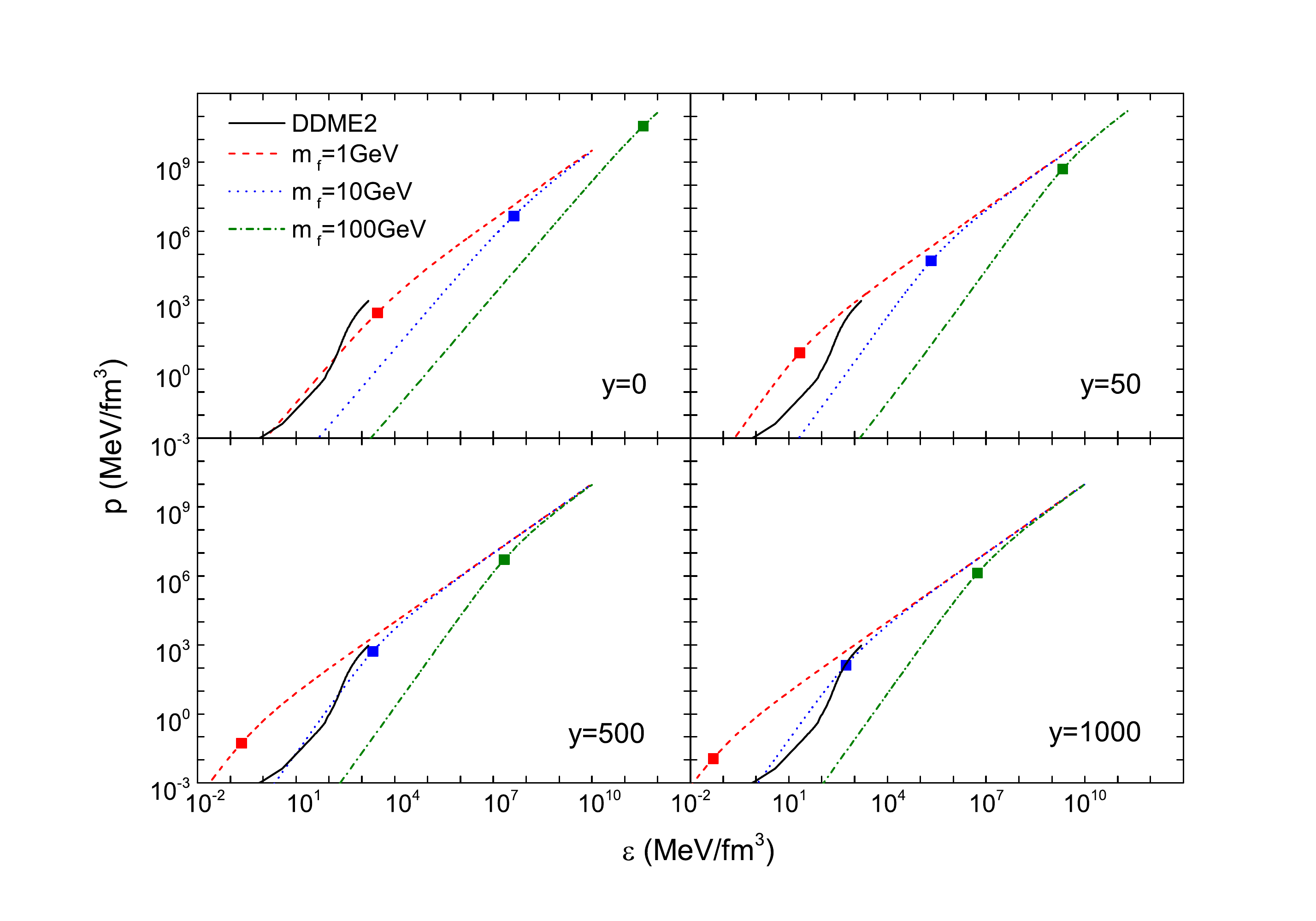}
    \caption{(color online). The EOSs for DM, namely the pressure as a function of energy density for the different particle mass of fermionic DM ($m_f$=1,~10,~100 GeV) and strength parameters ($y$=0,~50,~500,~1000).
        The EOS of NM from RMF with  parameter set DDME2 is shown for comparison. The squares on each line denote the central energy density where the mass of pure DM star takes the  maximum values.}
    \label{fig:eos} %% label for entire figure
\end{figure}

In Fig.~\ref{fig:eos}, we show the equation of states for DM, namely the pressure as a function of energy density,  for different $m_f$ (1, 10, 100 GeV) and $y$ (0, 50, 500, 1000). The EOS of NM from RMF with  parameter set DDME2 is shown for comparison. The squares on each line denote the central energy density where the mass of pure DM star takes the  maximum values.
As shown in Fig.~\ref{fig:eos}, when $y$ is fixed, the pressure is larger at the same  energy density of DM for the smaller $m_f$. However, when $m_f$ is fixed, the pressure is smaller at the same  energy density for the smaller $y$.

\begin{figure}[h!]
    \centering
    \includegraphics[height=5.5cm]{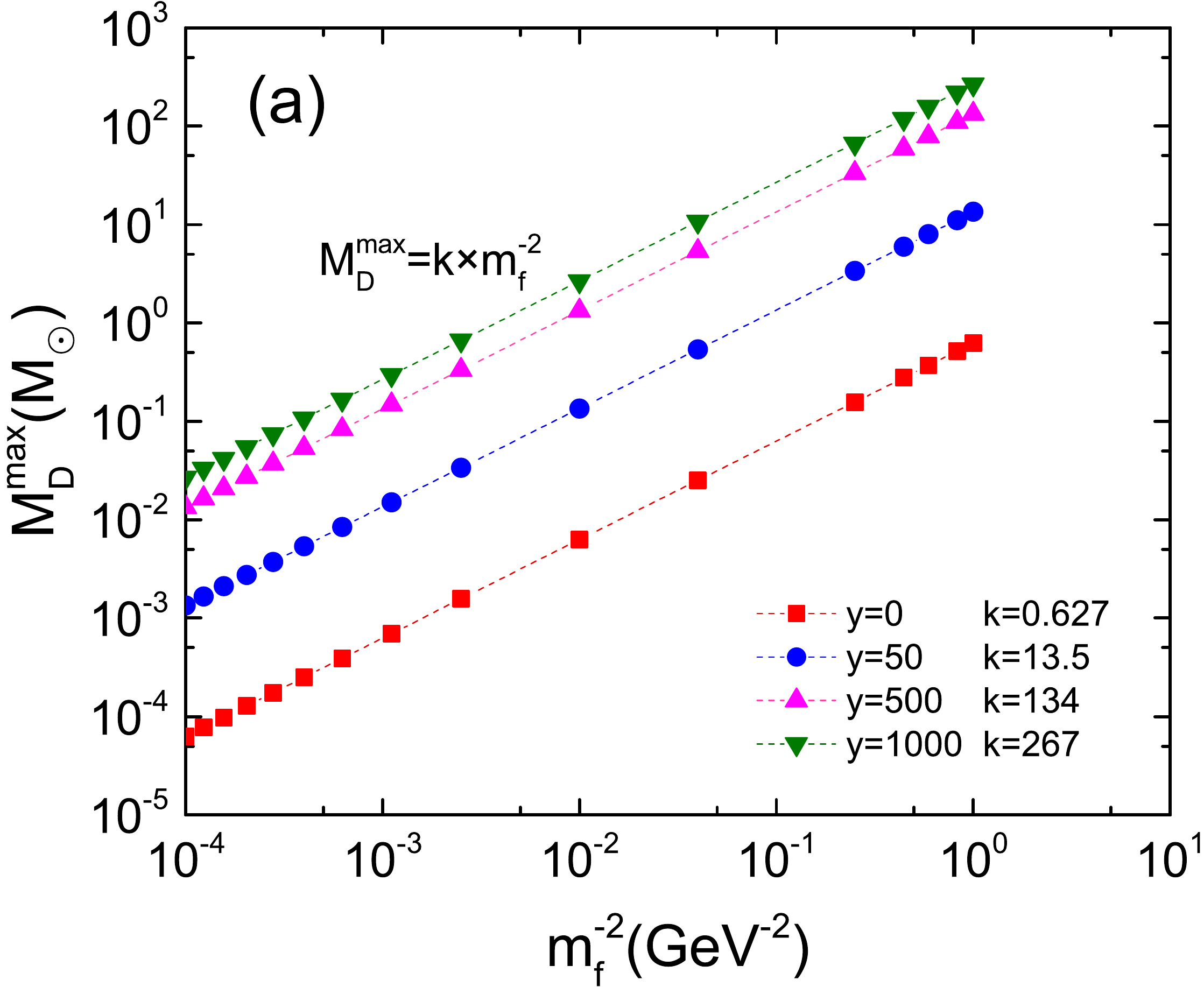}
    \includegraphics[height=5.5cm]{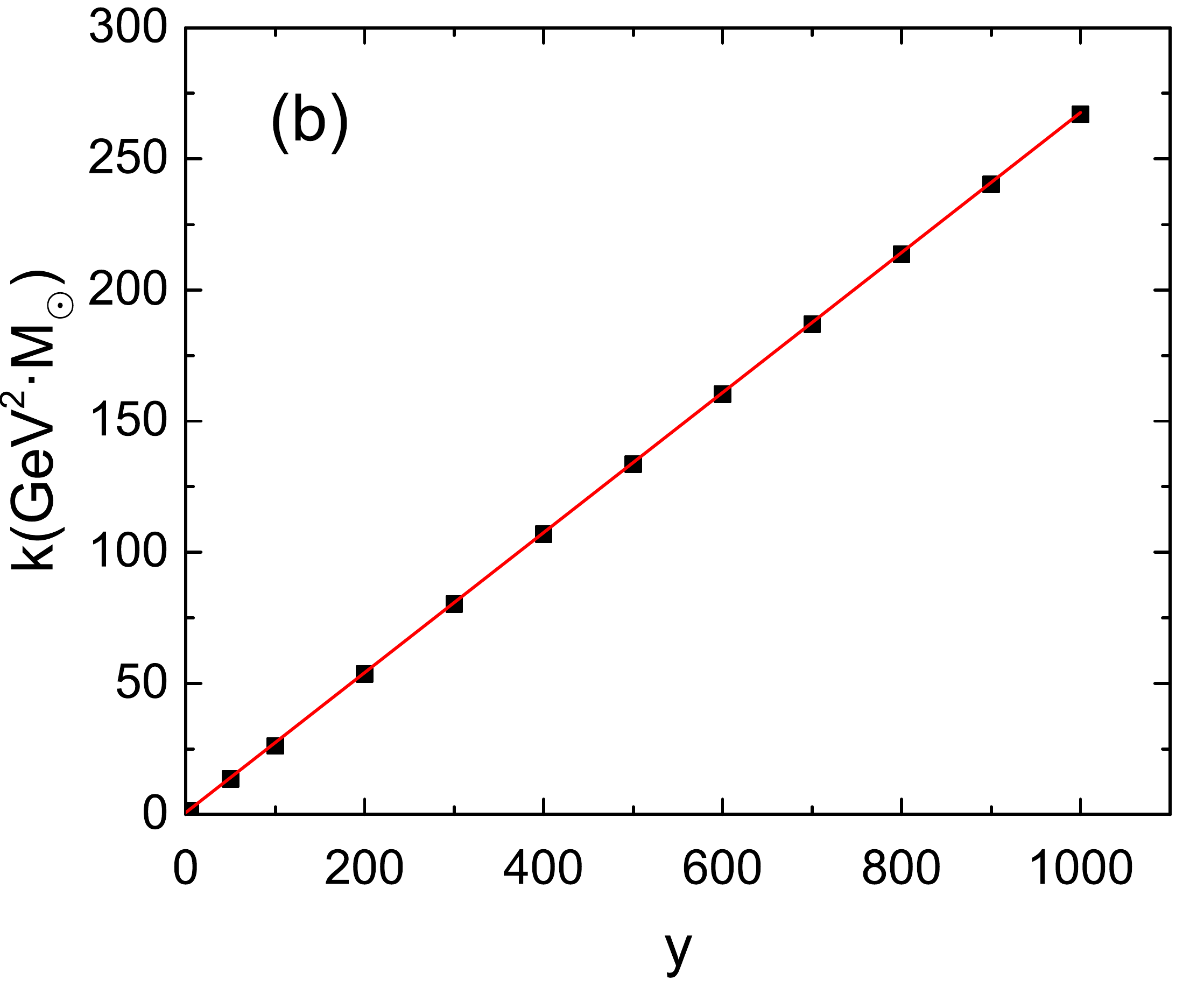}
    \caption{(color online).  The result for the pure DM stars.   Panel (a): The maximum mass of pure DM stars $M_{D}^{\max}$  as a function of $m_f^{-2}$ for different $y$ values. The fitted values of $k= M_{D}^{\max}\cdot m_f^{2}$ are listed in the figure.
        Panel (b): The relationship between the $k$ values  and  $y$. The points represent the present calculated results, and the line is the suggested relationship  $k=(0. 269y+0. 627)\gev^2\m_{\odot}$  given in Ref.~\cite{NarainPRD2006}. }
    \label{fig:DMSTAR}
\end{figure}

For the completion of discussion, we first calculate the properties of pure DM stars. The detailed discussions for the pure DM stars can be seen in Ref.~\cite{NarainPRD2006}.  The calculated maximum mass of pure DM stars $M_{D}^{\max}$   for different $m_f$ and $y$ are shown in Fig.~\ref{fig:DMSTAR}a. It is found that there is a linear relationship
between $M_{D}^{\max}$ and $m_f^{-2}$. The values of $k= M_{D}^{\max}\cdot m_f^{2}$ are almost constant for fixed strength parameters, i.e., $k$=0.627, 13.5, 134, 267 GeV$^2 \M_{\odot} $ for $y$=0, 50, 500, 1000, respectively. In Fig.~\ref{fig:DMSTAR}b, we list  more results of   $k$ values  for different $y$, which are denoted by the squares. These values are in consistent with the suggested relationship
\begin{equation}\label{eq:DMstar}
M_D^{\max}=(0.269 y+0.627)( \frac{1\GeV}{{m_f}})^2 \M_{\odot},
\end{equation}
given in Eq.(47) of Ref.~\cite{NarainPRD2006}, which is plotted by the line in Fig.~\ref{fig:DMSTAR}b. Such a simple relationship inspires us to search for the similar formula in DM admixed neutron stars (DANSs), which are shown in the following Figs. 3-7.

\begin{figure}[h!]
    \centering
    \includegraphics[width=7cm]{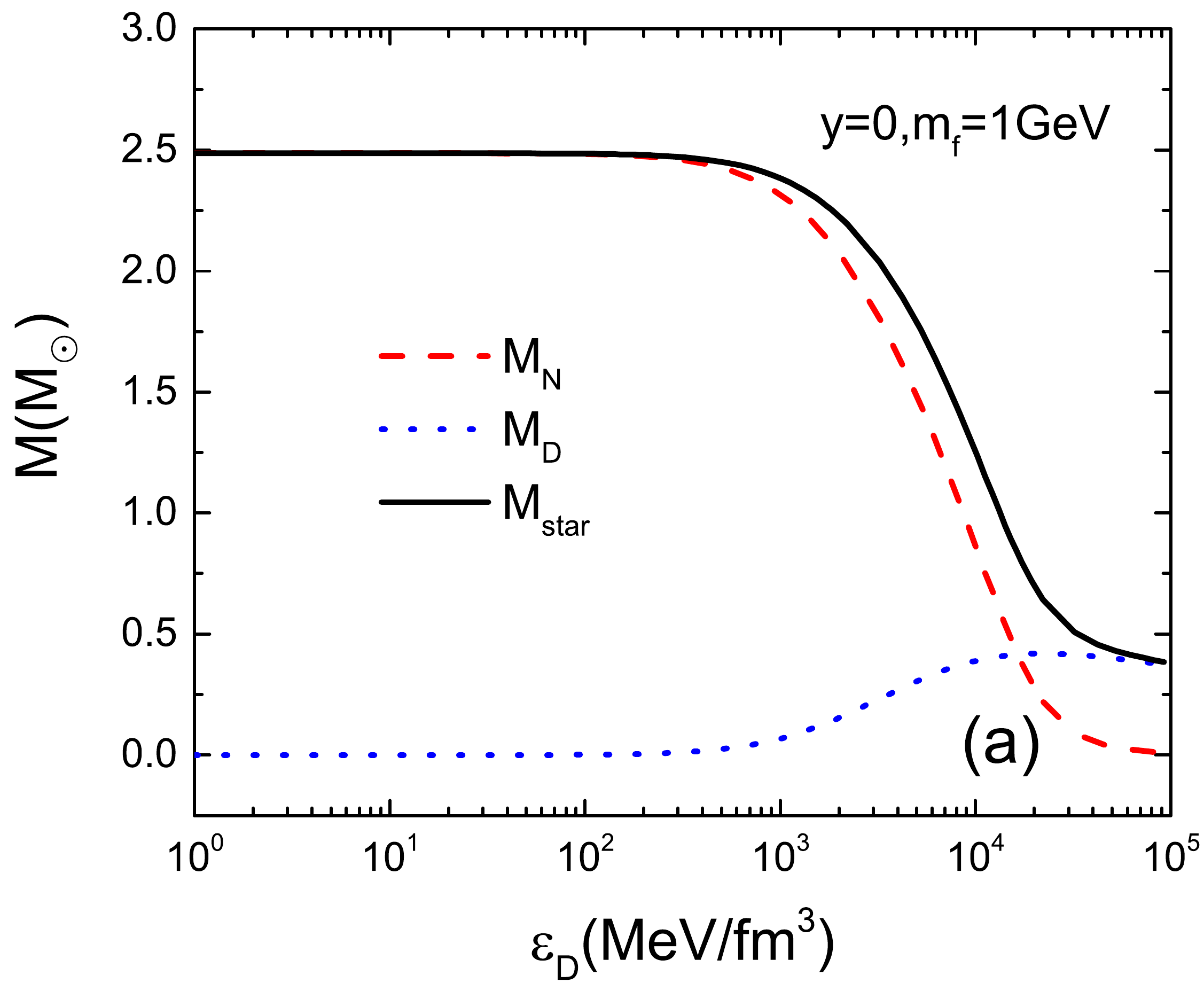}
    \includegraphics[width=7cm]{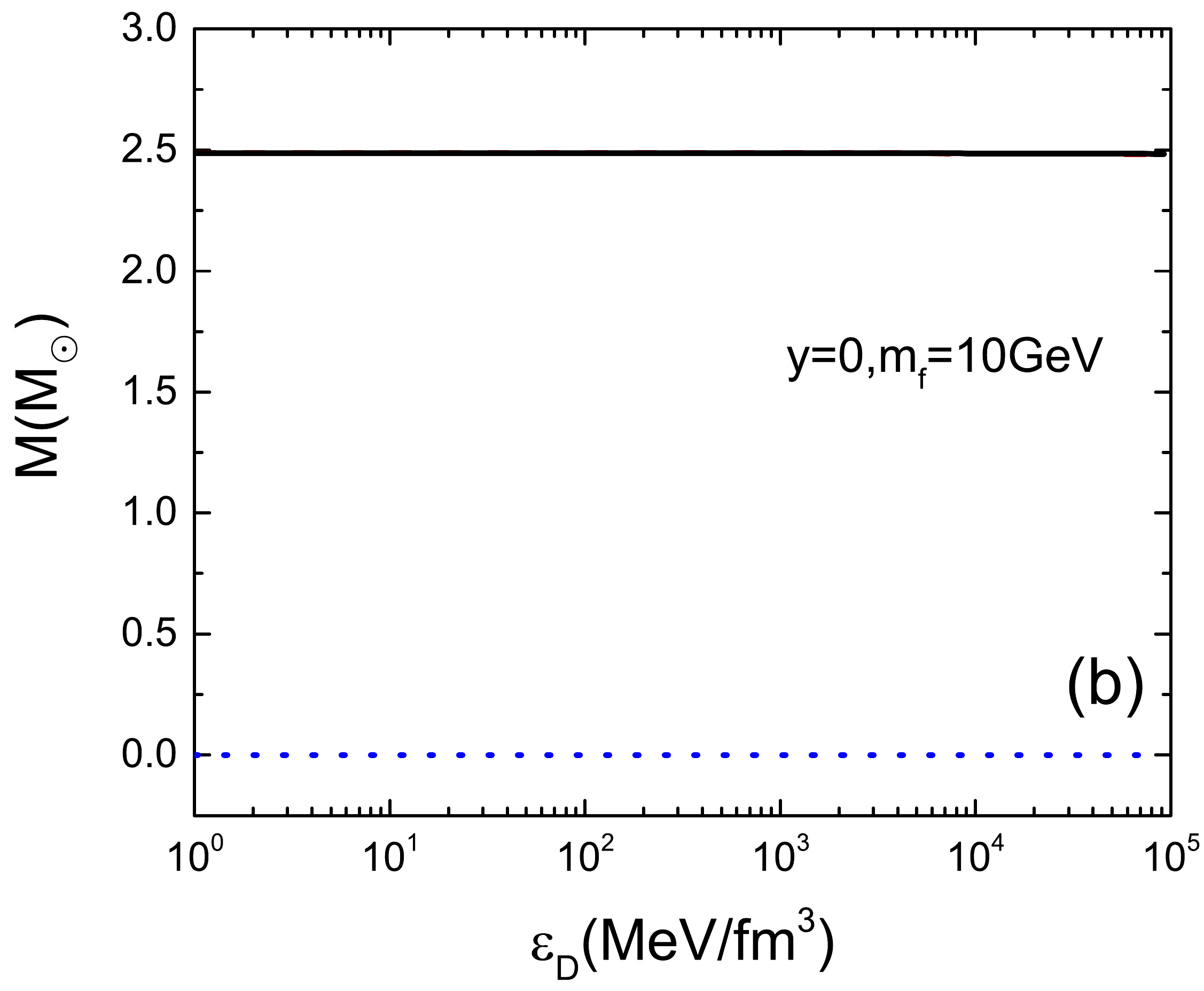}
    \includegraphics[width=7cm]{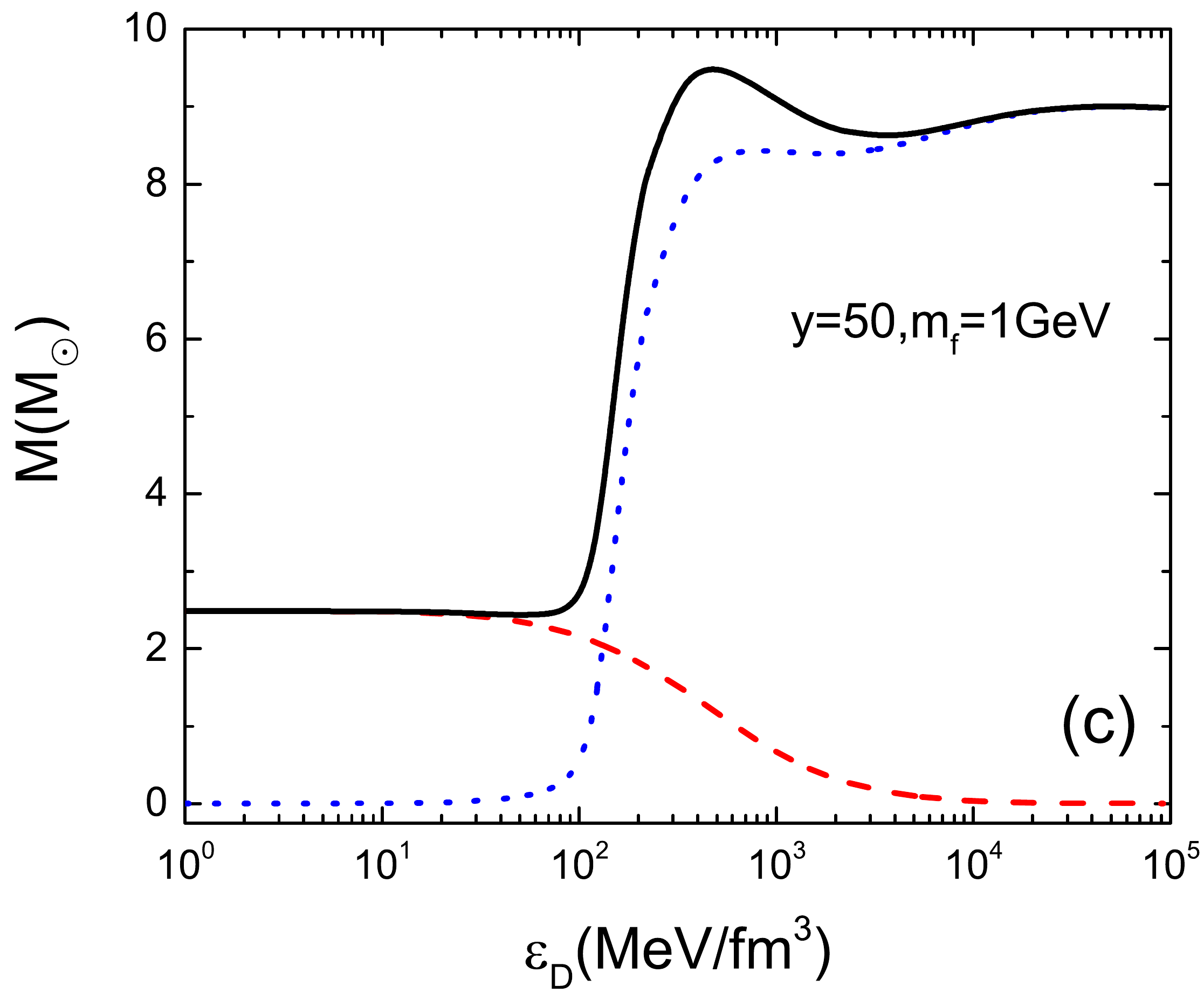}
    \includegraphics[width=7cm]{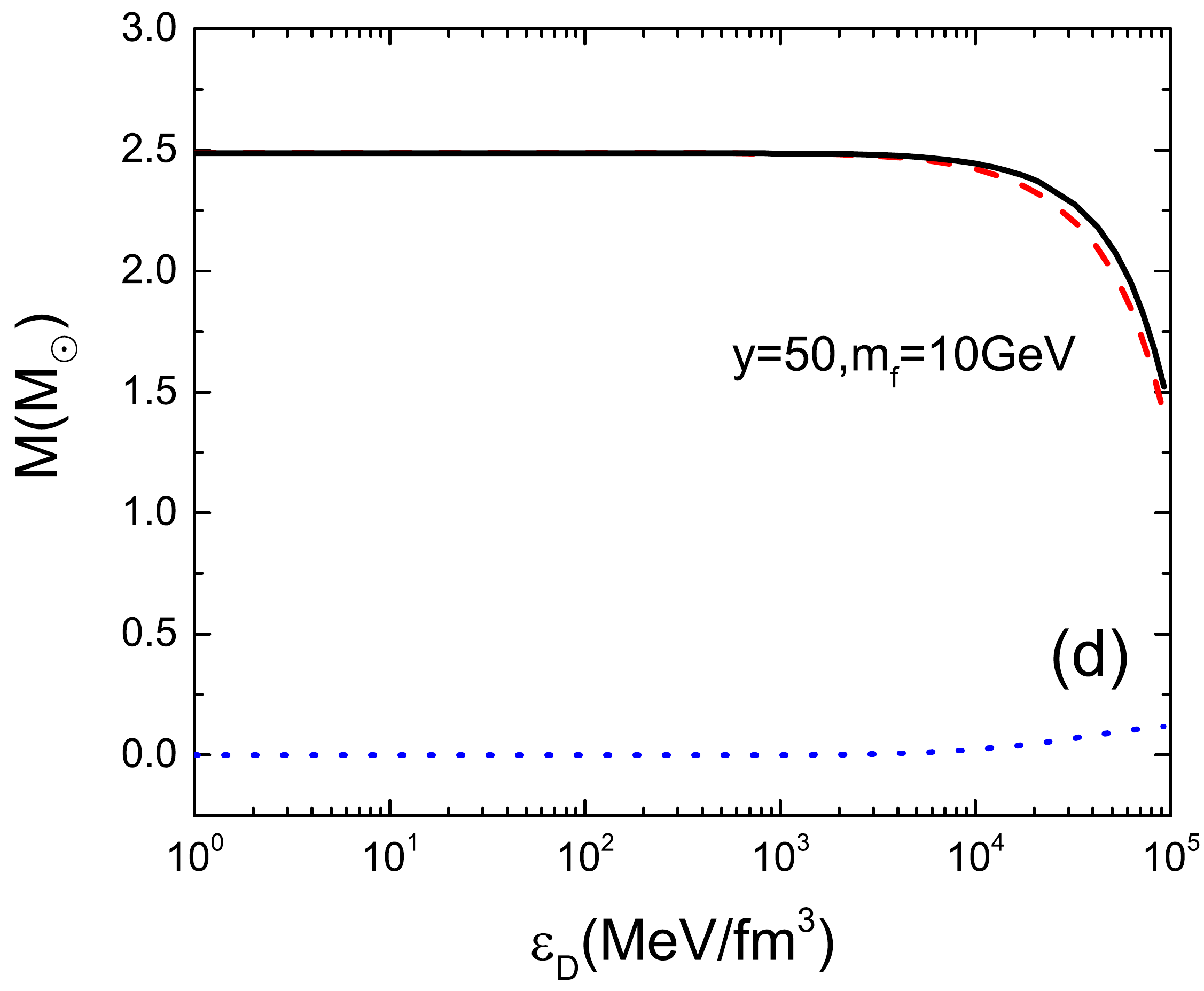}
    \hspace{0cm}
    \caption{(color online).  For DM admixed neutron stars,   the mass of DM $M_{D}$, the mass of NM $M_{N}$ and the total mass of compact star $M_{star}$  versus  the central energy density of DM $\varepsilon_D$. The EOS of NM is from RMF with DDME2 parameter set,  and the central energy density $\varepsilon_N$ is fixed as 1000MeV/$\fm^3$.
        The results for $m_f=1, 10$ GeV  and $y=0, 50$ of DM are shown.  }
    \label{fig:MDANS}
\end{figure}

For a DANS, we give the mass of the DM $M_{D}$, the mass of NM $M_{N}$ and the total mass of compact star $M_{star}$ in Fig.~\ref{fig:MDANS}, where the central energy density of NM $\varepsilon_N$ is fixed as 1000MeV/fm$^3$ and the central energy density of DM $\varepsilon_D$ varies. The NM is calculated based on the DDME2 EOS. As the examples, the results for $m_f=1, 10$ GeV  and $y=0, 50$ of DM are shown. The mass of neutron star without DM is 2.39$\M_\odot$ for DDME2 EOS when $\varepsilon_N$ is fixed as 1000MeV/fm$^3$.
As shown in Fig.~\ref{fig:MDANS}, $M_{N}$ decreases while $M_{D}$ increases with increasing of $\varepsilon_D$.
Such trend is more obvious when the $m_f$ is smaller and $y$ is bigger. For the case of  $m_f=1$ GeV  and $y=0$ as shown in the Fig.~\ref{fig:MDANS}a, $M_{D}$ is close to 0, and $M_{N}$, $M_{star}$ change little when $\varepsilon_D<$1000MeV/fm$^3$; then $M_{D}$ gradually increases to $0.4 \M_{\odot}$ and $M_{N}$ decrease to 0 when $\varepsilon_D>$1000MeV/fm$^3$.
For the case of $m_f=1$ GeV  and $y=50$ as shown in the Fig.~\ref{fig:MDANS}c, the $M_D$ increases obviously when $\varepsilon_D>$100MeV/fm$^3$, and $M_D$ could increase to 9.0 $\M_{\odot}$, while the $M_{star}$  increases up to 9.5$\M_{\odot}$.  If the particle mass of DM $m_f$ takes values of 10 GeV, the change of $M_{star}$ and  $M_{N}$ is no longer obvious. It is found  that the mass of DANSs usually decreases compared with the mass of neutron star without DM, but it could be tens or hundreds times of solar mass for small $m_f$ and large $y$.

\begin{figure}[h!]
    \centering
    \includegraphics[width=12cm]{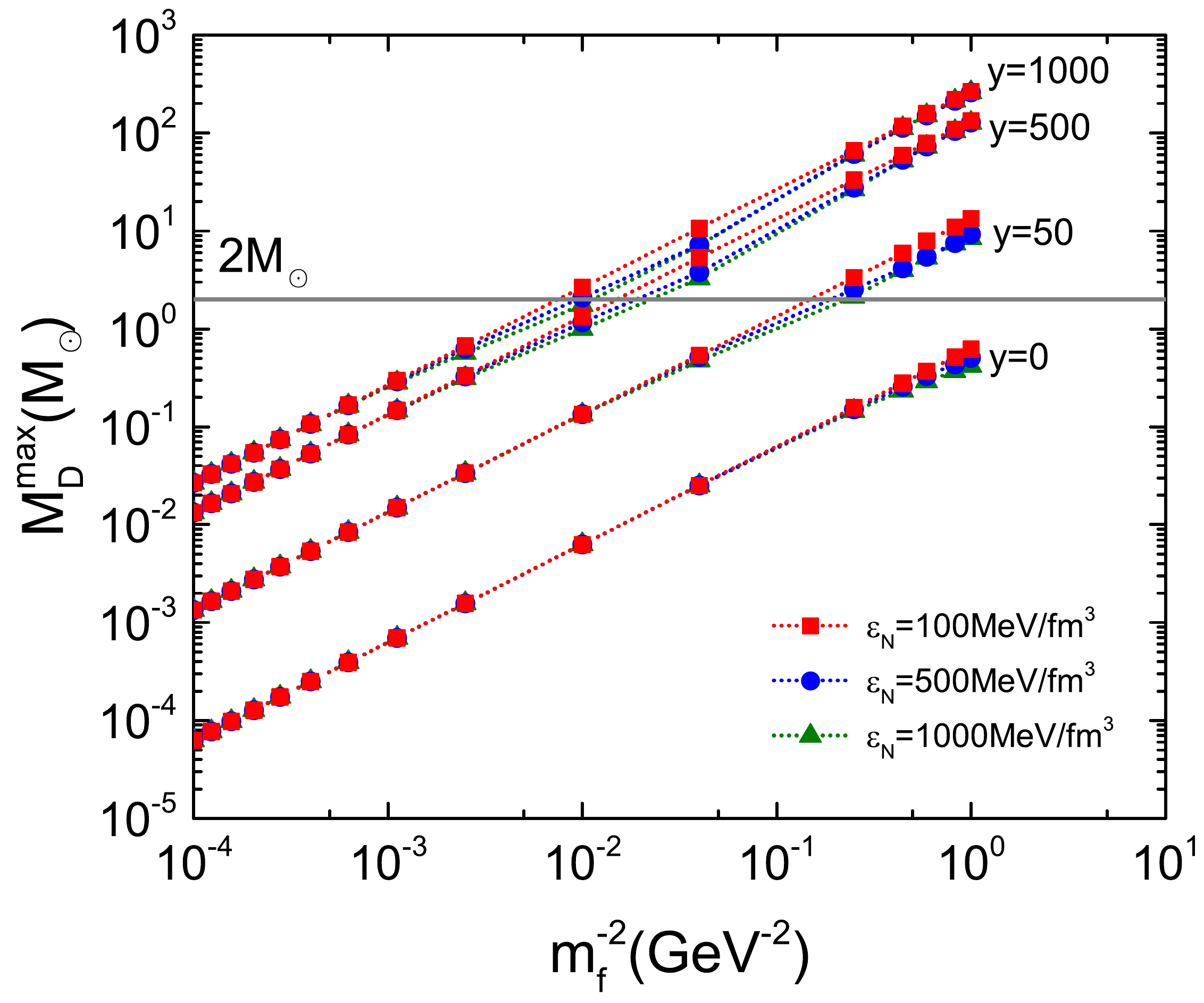}
    \caption{(color online). The maximum mass of DM $M_{D}^{\max}$ existing in DANSs as a function of $m_f^{-2}$ for different $y$ values.  The EOS of NM is from RMF with DDME2 parameter set, and the central energy density $\varepsilon_N$ is fixed as 100, 500 and 1000 MeV/fm$^{3}$, respectively. }
    \label{fig:m-mf} %% label for entire figure
\end{figure}

In the next step, we explore the  relationship between $M_D^{\max}$ (the maximum mass of DM in the DANSs) and the properties of fermionic DM, namely $m_f$ and $y$.
In Fig.~\ref{fig:m-mf},  $M_{D}^{\max}$ as a function of $m_f^{-2}$ for different strength parameters $y$ is given. The horizontal line represents the observed maximum mass $\sim 2.0$ M$_\odot$ of neutron star~\cite{DemorestNature2010, AntoniadisScience2013}. The central energy density of NM $\varepsilon_N$ is fixed as 100, 500 and 1000 MeV/fm$^{3}$, respectively.  The star mass of the neutron star without DM  will be 0.188, 1.957, and  2.390 $\M_{\odot}$ for $\varepsilon_N$=100, 500, 1000 MeV/fm$^{3}$, respectively. It is interesting to find that $M_{D}^{\max}$ has a liner relationship with $m_f^{-2}$ for DANSs, i.e., $M_{D}^{\max}=k\cdot m_f^{-2}$ for fixed $y$. It is amazing to find  that the coefficient $k$ is approximately independent of  $\varepsilon_N$ of NM.

\begin{figure}[h!]
    \centering
    \includegraphics[width=7cm]{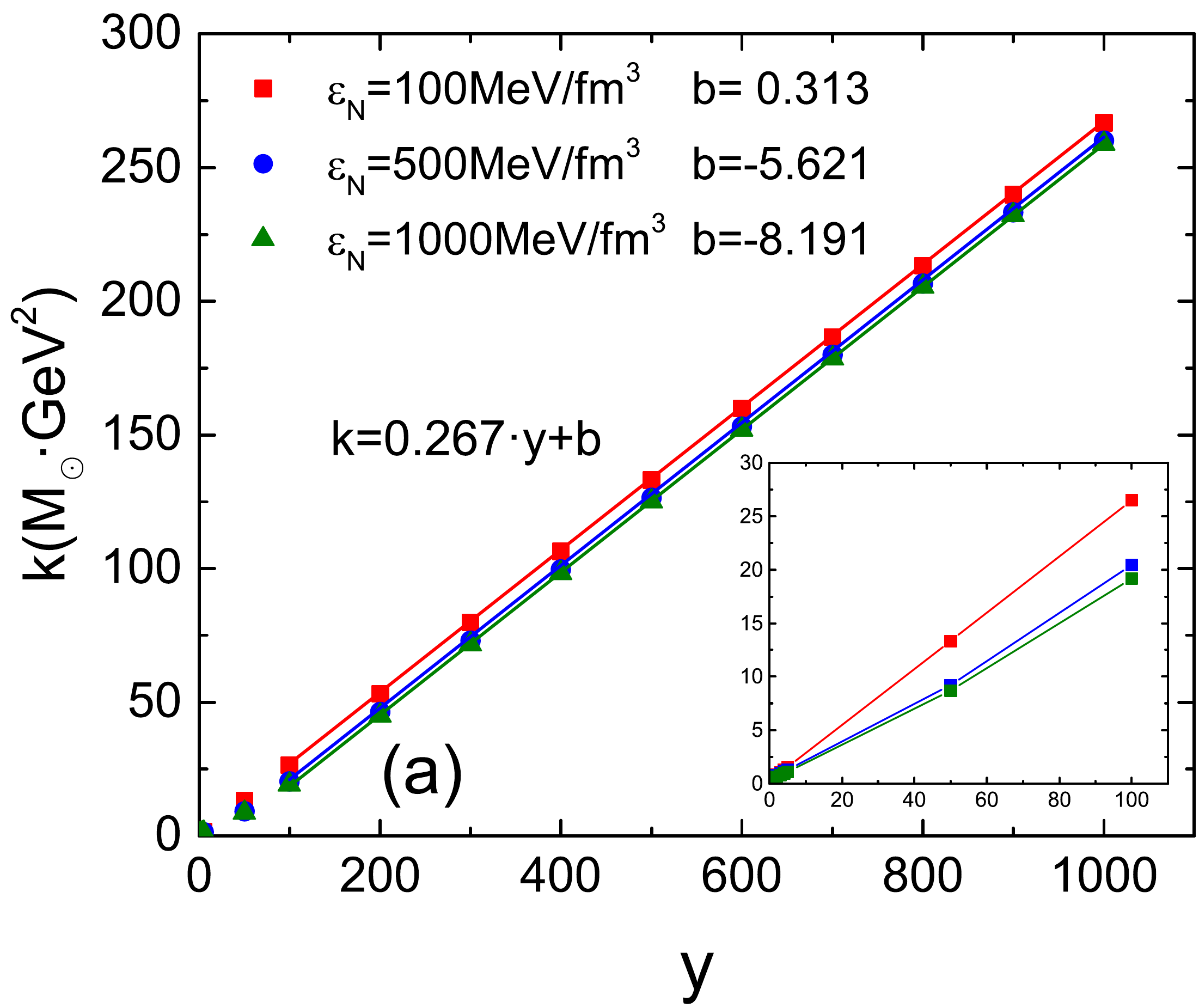}
    \includegraphics[width=7cm]{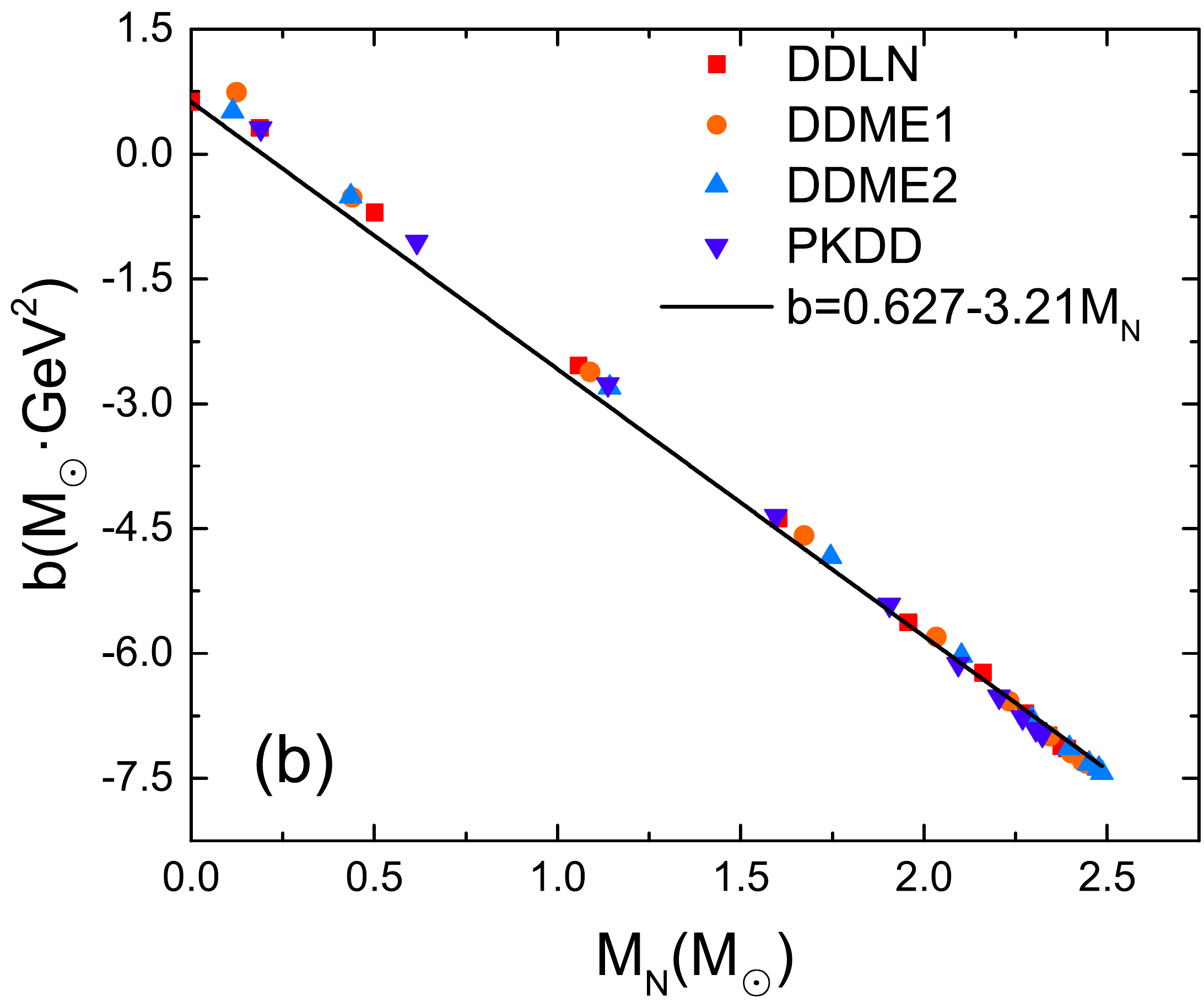}
    \caption{(color online).  Panel (a): $k=M_{D}^{\max}\cdot m_f^{2}$ as a function of $y$, where $M_{D}^{\max}$ is the maximum mass of DM in DANSs.  The EOS of NM is from RMF with DDME2 parameter set, and the central energy density $\varepsilon_N$ is fixed as 100, 500 and 1000 MeV/fm$^{3}$.
        The insert panel shows the relationship of $y<100$. The lines denote fitted values $k=0.267y+b$. The intercepts $b$ are shown for each line. Panel (b):
        The relationship of intercepts $b$ and $M_N$ (the mass of compact star without DM)  for different RMF EOSs with DDLN, DDME1,DDME2,PKDD parameter sets. The lines denote fitted values $b=0.627-3.21M_N$.   }
    \label{fig:k-y} %% label for entire figure
\end{figure}

To express $k$ in detail,  the relationship between the parameter $k$ and the strength parameter $y$  with different $\varepsilon_N$  is fitted in the Fig.~\ref{fig:k-y}.
It is found that the fitted relationship of $k=0.267\cdot y+b$ is in good agreement with the calculated $k$ values for $y>100$.  For different $\varepsilon_N$, the values of $b$ are different, e.g., 0.313($\varepsilon_N$=100MeV/fm$^3$), -5.621($\varepsilon_N$=500MeV/fm$^3$) and -8.191($\varepsilon_N$=1000MeV/fm$^3$). In further, we found it is  a linear relationship between $b$ and  and $M_N$ (the mass of neutron star without DM), namely $b=0.627-3.21 M_N$. The relationship $k=0.267y+0.627-3.21M_N$ is independent of the different RMF EOSs with DDLN, DDME1, DDME2, PKDD parameter sets adopted in the calculations.

In a word, the  relationship between $M_D^{\max}$ (the maximum mass of DM in the DANSs) and the properties of fermionic DM can be suggested as
\begin{equation}\label{eq:M-mf,y2}
M_D^{\max}=(0.267 y +0.627-3.21\frac{M_N}{\M_{\odot}})( \frac{1\GeV}{{m_f}})^2 \M_{\odot}.
\end{equation}
When $M_N=0$, the relationship is in consist with above mentioned formula as suggested for pure DM star in  Ref.~\cite{NarainPRD2006}.
In Ref.~\cite{NarainPRD2006}, it is pointed that the relationship of  pure DM star  is non-linear  for $y<1$, here  the relationship for DANSs is non-linear for $y<100$, as shown in the insert plot of Fig.~\ref{fig:k-y}a.

\begin{figure}[h]
    \centering
    % label for first subfigure
    \includegraphics[width=12cm]{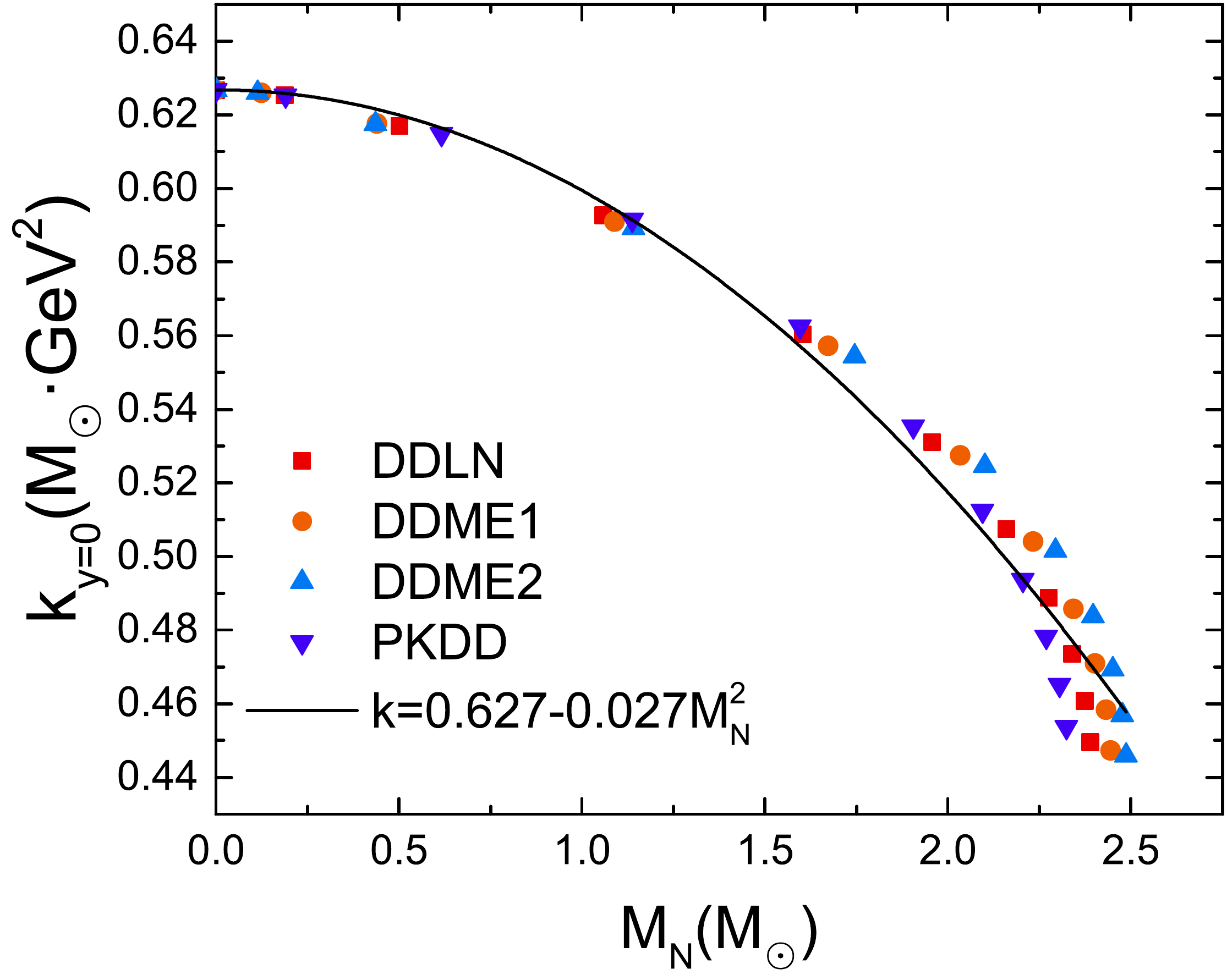}
    \hspace{0cm}
    \caption{(color online).  For free fermion DM model ($y$=0), the values of $k=M_{D}^{\max}\cdot m_f^{2}$ as a function of $M_N$, where $M_{D}^{\max}$ is the maximum mass of DM in DANSs, $M_N$ is the mass of compact star without DM. The results adopted different RMF EOSs with DDLN, DDME1,DDME2,PKDD parameter sets are shown. The line denotes the fitted relationship $k_{y=0}=0.627-0.027M_N^2$. }
    \label{fig:k0-mn}
\end{figure}

As a specific case in the non-linear region, $y=0$ corresponds to the free fermi DM model. For free fermi DM model, the values of $k_{y=0}=M_{D}^{\max}\cdot m_f^{2}$ as a function of $M_N$ are given in Fig.~\ref{fig:k0-mn}, where $M_{D}^{\max}$ is the maximum mass of DM in DANSs, $M_N$ is the mass of neutron star without DM. The results with different RMF EOSs with DDLN, DDME1, DDME2, PKDD parameter sets are shown. We can see clearly that the relationship between $k_{y=0}$ and $M_N$ for all the adopted RMF EOSs can be approximatively described by a parabola equation, which is fitted as $k_{y=0}=0.627-0.027M_N^2$ and shown as the solid line in Fig.~\ref{fig:k0-mn}. This decreasing tendency indicates that the dependence between $M_{D}^{\max}$ and $m_f^{-2}$ is suppressed with increasing $M_N$. If we set $M_N=0$, then we go back to the pure DM stars with $y=0$ and $k=0.627$.

\begin{figure}[h]
    \centering
    \label{fig:rd-mdmax}
    \includegraphics[width=12cm]{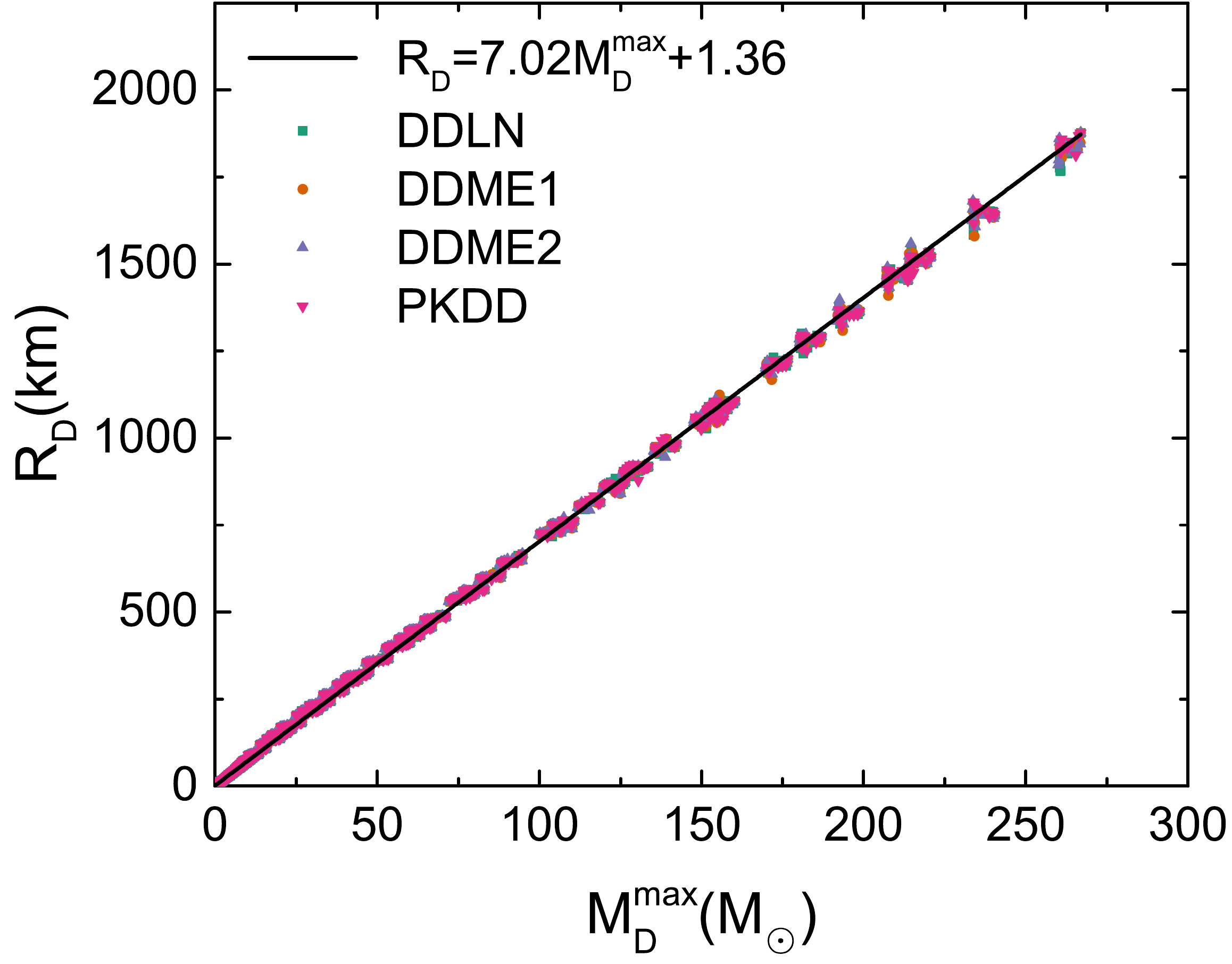}
    \caption{(color online). The relation between radius of DM and the mass of DM in DANSs when the mass of DM in DANSs is maximal.  The symbols denote the 10880 sets of results for different NM EOSs. The line denotes the fitted relationship ${R_D}
        =7.02 M_D^{\max}+1.36$. }
\end{figure}

Similar to the mass, radius is also one of the most important observable quantity for compact star and should be studied in detail. For pure DM stars, Ref.~\cite{NarainPRD2006} has suggested a equation for radius where maximum mass is obtained based on $y$ and $m_f$.  The relationship between radius of DM and the mass of DM in DANS when the mass of DM is maximal is shown in Fig.~7. As mentioned above, we adopt 16 values for  $m_f$ and 17 values for $y$. In addition,  the central energy density of nuclear matter $\varepsilon_N$ takes values of  $100$,  $200$, $300$, $400$, $500$, $600$, $700$, $800$, $900$, and $1000$ MeV$/$fm$^{3}$. The same calculations are performed for DDLN, DDME1, DDME2, PKDD EOSs and thus total 10880 points are obtained. A universal relationship is found to exist for all the points in the $R_D-M_D^{\max}$ plane. The corresponding relationship can be precisely described by
\begin{equation}
R_D=\left(7.02\cdot \frac{M_D^{\max}}{ M_{\odot}}+1.36\right) \km.
\end{equation}

\section{Conclusion}

By assuming that only gravitation acts between dark matter (DM) and normal matter (NM), we studied DM admixed neutron stars (DANSs) using the TOV equations for two-fluid separately.
The NM and DM of compact stars are simulated by the relativistic mean field (RMF) theory and  non-self-annihilating self-interacting fermionic model, respectively.
The effects of the mass of DM fermion $m_f$ and the interaction strength parameter $y$  on the properties of DANSs are investigated in detail. Due to the lack of information about the particle nature of DM, we consider the particle mass of fermionic DM $m_f$ and  the strength parameter $y$ as the free parameters.
It is found   that the mass of DANSs usually decreases compared with the mass of neutron star without DM, but it could be tens or hundreds times of solar mass for small $m_f$ and large $y$. For a DANS,  we suggest a universal relationship  $ M_D^{\max}=(0.267 y +0.627-3.21\frac{M_N}{\M_{\odot}})( \frac{1\GeV}{{m_f}})^2 \M_{\odot} $ for  $y>100$, where $ M_D^{\max} $ is the maximum mass of DM in DANSs  and $M_N$ is the mass of the neutron star without DM. For free fermion DM model ($y$=0),  the relationship becomes $ M_D^{\max}=(0.627-0.027\frac{M_N^2}{\M_{\odot}^2}) ( \frac{1\GeV}{{m_f}})^2 \m_{\odot}$.
The radius of DM $R_D$ shows a linear relationship with $M_D^{\max}$ in DANSs, namely $R_D
=(7.02\frac{M_D^{\max}}{ \M_{\odot}}+1.36)$km.
These conclusions are independent of the different NM EOSs from RMF theory. Such a simple universal relationship  connecting the nature of DM particle and mass of stars  might shed light on the constraining the nature of the DM by indirect method.

\section*{Acknowledgments}
    This work is partly supported by the National Natural Science Foundation of China (Grant Nos. 11675094, 11622540), and Young Scholars Program of Shandong University, Weihai (Grant No. 2015WHWLJH01).

%\section*{References}

\end{document}